# Classification of red blood cell shapes in flow using outlier tolerant machine learning

Alexander Kihm[1], Lars Kaestner[1,2] Christian Wagner[1,3,*] Stephan Quint[1]

1 Department of Experimental Physics, Saarland University, Campus E2 6, 66123 Saarbrücken, Germany
2 Theoretical Medicine and Biosciences, Saarland University, Campus University Hospital, Building 61.4, 66421 Homburg, Germany
3 Physics and Materials Science Research Unit, University of Luxembourg, Luxembourg

*c.wagner@uni-saarland.de

## Abstract

The manual evaluation, classification and counting of biological objects demands for an enormous expenditure of time and subjective human input may be a source of error. Investigating the shape of red blood cells (RBCs) in microcapillary Poiseuille flow, we overcome this drawback by introducing a convolutional neural regression network for an automatic, outlier tolerant shape classification. From our experiments we expect two stable geometries: the so-called 'slipper' and 'croissant' shapes depending on the prevailing flow conditions and the cell-intrinsic parameters. Whereas croissants mostly occur at low shear rates, slippers evolve at higher flow velocities. With our method, we are able to find the transition point between both 'phases' of stable shapes which is of high interest to ensuing theoretical studies and numerical simulations. Using statistically based thresholds, from our data, we obtain so-called phase diagrams which are compared to manual evaluations. Prospectively, our concept allows us to perform objective analyses of measurements for a variety of flow conditions and to receive comparable results. Moreover, the proposed procedure enables unbiased studies on the influence of drugs on flow properties of single RBCs and the resulting macroscopic change of the flow behavior of whole blood.

## Author summary

Artificial neural networks represent a state-of-the art technique in many branches of natural sciences due to their ability to fastly detect and categorize image features with high throughput. We use a special type of neural network, the so-called convolutional neural network (CNN) for the classification of human red blood cell shapes in microcapillary Poiseuille flow. Following this approach, phase diagrams of two distinct classes (slippers, croissants) are generated and, by comparison with a manually obtained phase diagram, optimized threshold ranges for categorizing the output values are established. This allows us to better understand the complex fluid behavior of blood depending on the intrinsic properties of single red blood cells. For future studies, we aim to predict phase diagrams under the influence of certain drugs.

# Introduction

Amongst all human organs, blood is the most delocalized one, delivering oxygen from the respiratory system to the tissues in the body and transporting carbon dioxide back. On a microscopic scale, this is performed by red blood cells (RBCs) which form the largest fraction of cells in whole blood ($\approx 99\,\%$). At rest, RBCs are biconcave discocytes with an average diameter of $8\,\mu$m and a height of $2\,\mu$m. Due to their flexible membrane, RBCs alter their shape under external stress prevalent in the microvascular network [1,2]. This feature is one of the key properties of RBCs, which allows them to squeeze through geometrical constrictions much smaller than their stress-free shape [3], which is partly an intrinsic property of RBC morphology [4,5] and partly an active adaptation process [6,7]. Although data on the mechanical properties of RBC suspensions is widely known from rheological measurements [8], the linkage to individual cell behavior is limited. Further, the comparison between capillary Poiseuille flow and pure shear flow prevalent in rheometers is difficult. Consequently, mimicking flow under physiological conditions in vitro demands for experimental setups such as PDMS-based microchannels, ubiquitous in lab-on-a-chip devices [9]. In this work, we focus on experiments of individual flowing RBCs, providing a holistic insight to individual cell mechanics. Hereby, the experimental data originate from a previous study on RBC shape geometry [10] and the data is reused for the introduction of a fully automated data analysis approach based on a deep learning convolutional neural network (CNN) [11].

As described in the preceding work, two stable RBC shapes are expected from the measurements: The so-called 'croissant' and the 'slipper' shape [12–14]. Their frequency of occurrence highly depends on the imposed flow conditions. Whereas axisymmetric croissants mostly appear at lower flow velocities, non-axisymmetric slippers are observed at higher shear rates [15,16]. However, besides these stable geometries, also a large number of indefinite shapes occur, especially in the so called phase transition range. These outliers or 'other' shapes make a considerable amount of the whole statistics. Due to their large shape variance, these cells cannot be assigned to a mutual exclusive class and can therefore not easily be involved in the machine learning processes. Overcoming this drawback, we present a regression based CNN aiming to distinguish between croissants, slippers and others by applying statistically derived thresholds to the net response. As a result, we obtain an automatically generated RBC phase diagram which relinquishes any subjective user input. Prospectively, we will be able to run comparable studies on RBC shape geometries in microchannels of variable size and to generate unbiased phase diagrams. Above all, data evaluation can then be performed in a highly time-effective manner.

# Materials and Methods

## Ethics Statement

Human blood withdrawal from healthy donors as well as blood preparation and manipulation were performed according to regulations and protocols that were approved by the ethic commission of the "Ärztekammer des Saarlandes" (reference No 24/12). We obtained informed consent from the donors after the nature and possible consequences of the studies were explained.

## Architecture of convolutional neural networks

CNNs are digital image processing systems [17] with the ability to resolve and evaluate the details of an input image. Usually, they are used to e.g. recognize and classify particular objects or humanoid faces [18] within pictures. Here, a CNN is exploited to distinguish between the shape characteristics of RBCs in flow and to detect undefined outliers which occur due to channel imperfections, membrane damages, cell-cell-interactions, shape transitions or transients, and optical ambiguities.

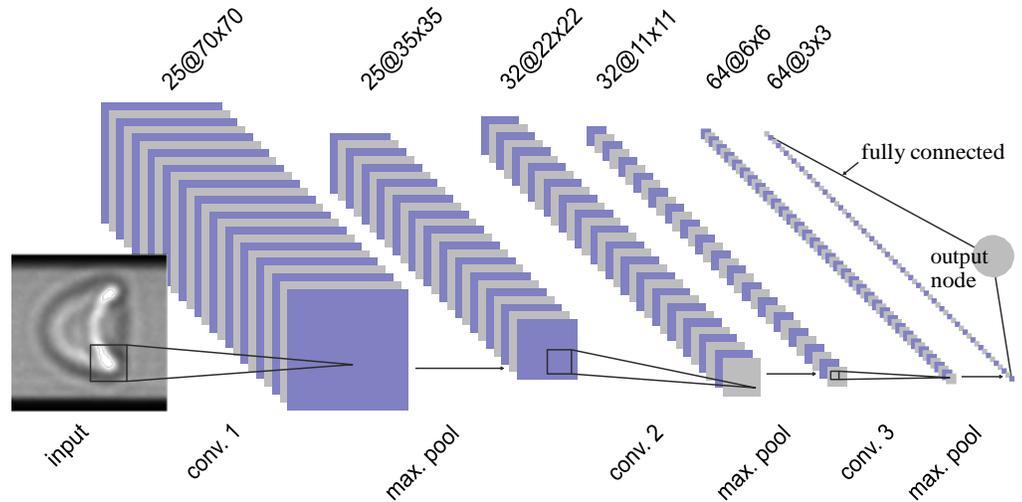

**Fig 1.** Layer structure of the used CNN. The input layer accepts cell images of $90 \times 90\,\text{px}^2$. Avoiding border effects (top and bottom) caused by irregular light refraction at the channel edges, input images are weighted by a Tukey window ($\alpha = 0.25$) causing a fading effect towards the upper and lower edge (Eq. 3). In a first processing stage, images are convoluted by 25 different convolution kernels of size $21 \times 21$. This results in 25 intermediate images of size $70 \times 70\,\text{px}^2$, which undergo a non-linear rectification (reLU layer) before getting down-sampled by a max-pooling procedure ($2 \times 2$, stride 2) to a size of $35 \times 35\,\text{px}^2$. The combination of convolution, rectification, and max-pooling are repeated twice using different sets of convolution kernels (see Tab. 1). The output node then intertwines all resulting subimages by a full interconnection of all available pixel values and maps them to a linear output range. Sizes of subimages (blue/grey) as well as the indications of convolution kernels (black) are chosen to scale, illustrating that kernels obey the characteristic features of input and subimages.

Independent of the particular use case, the architectures of CNNs usually follow the same design rules. They consist of an image input layer followed by a certain number of subsequent convolution stages (Fig. 1), and provide so-called interconnected layers forming an artificial neural network (ANN) to combine the convolution data in a final stage before the information is fed to the output layer nodes.

The main stages of a CNN usually consist of several sublayers [19] including the actual convolutional layer, a non-linear rectification 'reLU' layer [20], and a pooling layer [21]. Even though more sophisticated designs [22, 23] exist, for RBC shape recognition we restrict ourselves to these layer types keeping the system as simple as possible.

Convolutional layers make use of a number of convolutional kernels (feature maps) of particular size and are optimized to find the major characteristics of a set of various input images being subject to a sophisticated training process (cf. section Training). For instance, this can include horizontal or vertical edges but also more detailed characteristics, such as the 'tail' of a cell, see Fig. 2. Convolving an image with a set of specially optimized convolution kernels then results in a number of output images showing clear differences in between the RBC shape classes.

| layer | kernel size [px$^2$] | subimage size [px$^2$] |
|---|---|---|
| input layer | - | 90 × 90 |
| conv. layer 1 | 21 × 21 | 70 × 70 |
| reLU layer | - | 70 × 70 |
| max pooling layer | 2 × 2, stride 2 | 35 × 35 |
| conv. layer 2 | 14 × 14 | 22 × 22 |
| reLU layer | - | 22 × 22 |
| max pooling layer | 2 × 2, stride 2 | 11 × 11 |
| conv. layer 3 | 6 × 6 | 6 × 6 |
| reLU layer | - | 6 × 6 |
| max pooling layer | 2 × 2, stride 2 | 3 × 3 |
| output layer, regression type | - | 1 × 1 |

**Table 1.** Overview of the used layers in the indicated deep learning CNN. In the center column, the kernel size of the corresponding layer is given. The resulting image size after layer passage is given in the rightmost column.

In a mathematical sense, convolutions are linear operations. Thus, CNNs could easily be contracted to a simple linear signal processing system if there were no non-linearities involved. Indeed, the usage of non-linearities resolving higher order dependencies in between the characteristics of a set of distinct input images is one of the key ideas behind neural networks. For CNNs, it is common practice to use a reLU layer to set the negative values of the subimages to zero. This renders image transformations non-unitary and allows to introduce further non-linear evalution branches within the tree-like structure of the system.

After the convolution and rectification stage, the amount of data is fairly increased and usually corresponds to a multiple of the input data. Thus, it is reasonable to reduce data with the aid of pooling layers [21, 24]. Within such layer, subsets of each intermediate image are pooled according to their mean or maximum values. Results are then assigned to a target image of smaller size. For our CNN, subsets of size 2 × 2 px$^2$ with a stride of 2 px are evaluated according to the maximum value aiming to halve the sizes of intermediate images at the end of each convolution stage. A further advantage of such pooling strategy is to obtain certain tolerance with respect to spatial translations of objects.

As special feature, the here employed CNN makes use of a regression output layer providing a linear transfer function. In contrast to standard classification networks offering a purely binary output, realized e.g. by a logistic transfer function or softmax approach [25], the output node of our CNN is able to take on a range of floating point values. Taking the number space of input images into account (8-bit space), we define our output to move in the same range thus the same order of magnitude. Taking values at this scale, we observe a fast convergence of the training process. Consequently, perfect slippers are defined at value −127 whereas croissants are located at 127. Input images leading to an output value which significantly differs from these targets are assumed to be indefinite and therefore discriminated to be an outlier of type other (cf. section Training).

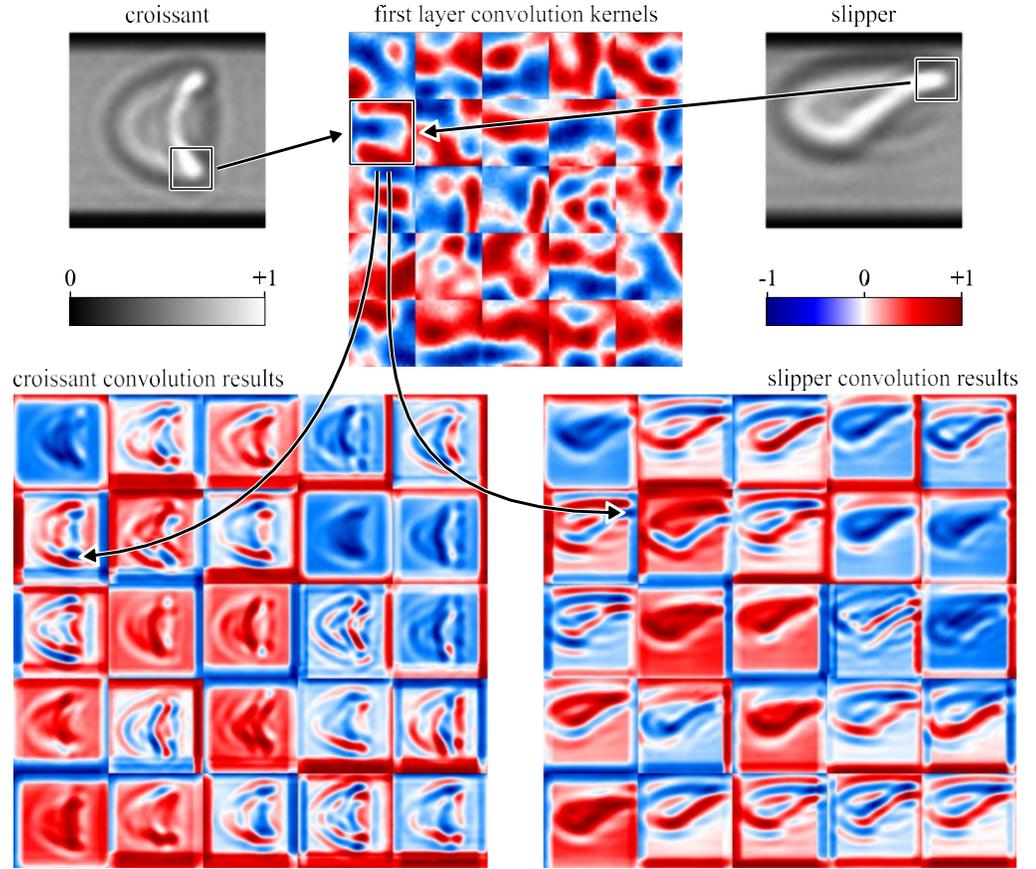

**Fig 2.** Resulting subimages (bottom) of two contrary RBC shapes (croissant, upper left; slipper, upper right) passing the first convolutional layer of a CNN. The convolution kernels as well as the subimages are represented by a false color mapping for the sake of better visibility. Boxes in the input images indicate typical features of both cell shape classes and the respective enhancement of these after convolution (indicated by arrows).

## Training of convolutional neural networks

CNNs are not ready to use if neurons are not 'trained' to solve certain regression or classification problems. This means that initially randomly chosen convolution kernels, weights and bias terms (concluded in the vector $\sigma$ of free parameters) first need to be reasonably updated. For this purpose, we employ a supervised learning approach which requires three major ingredients: A labeled set of training data, a suitable loss function that has to be minimized, as well as an appropriate optimization strategy.

As loss function, the root mean square error (RMSE) is chosen (Eq. 1), expressing the cumulated differences between all target and actual output values of the training data set.

$$\text{RMSE} = \left(\frac{1}{N}\sum_{i=0}^{N-1}(e_i)^2\right)^{1/2} = \left(\frac{1}{N}\sum_{i=0}^{N-1}(t_i - a_i)^2\right)^{1/2} \quad (1)$$

Here, $a_i$ expresses the actual output of a given input image $i$ and $t_i$ the respective, predefined target value.

Finding the global minimum of the RMSE, $\sigma$ is optimized using a stochastic gradient descent solver with momentum (SGDM), see [26] and [27]. The SGDM algorithm updates the vector $\sigma_l$ in discrete steps, where $l$ denotes the actual iteration:

$$\sigma_{l+1} = \sigma_l - \alpha \nabla E(\sigma_l) + \gamma (\sigma_{l+1} - \sigma_l). \qquad (2)$$

$E(\sigma_l)$ corresponds to the loss function, respectively the RMSE, $\alpha = 0.001$ to the constant learning rate, and $\gamma = 0.9$ to the momentum term.

Achieving a faster convergence, a single optimization iteration takes into account a so-called mini-batch [25, 28, 29] consisting of a subset (128 images) of the randomly shuffled training data. In this context, a full pass of the training data is called an 'epoch'. A typical temporal evolution of training states is shown in Fig. 3 where the loss function is plotted with respect to the number of training epochs. We set a maximum of ten training epochs, since a prolongation to more training epochs rather causes overtraining instead of a gain in performance. This overtraining of the neural network can be monitored in the validation loss: A termination criterion of the training status is implemented by a consecutive increase in validation loss five times, since a divergence between training and validation loss is an indicator for not being in the global optimum. Further, we choose a validation frequency (update frequency of validation loss) of 50 mini-batches, to not meet the termination criterion by random fluctuations of the training process. In general, the validation data set consists of data disjoint from the ones for training to guarantee an independent validation of the networks' performance. For our purposes, 5% of the training data set was chosen randomly for validation, whereas the rest was the true training data.

For training, besides slippers and croissants, we additionally define an auxiliary class called 'sheared croissant' (Fig. 4). This is due to the fact that the characteristic features of pure croissants and sheared croissants show a larger mutual similarity than pure croissants and slippers. Assuming a linear scale of output values, this subtype is biased towards a value of 64. Contrasting the nomenclature, they are not any type of croissants, but most probably resemble a slipper flipped by 90° perpendicular to the optical axis. However, due to the lack of information of our 2D images, we cannot make a clear statement on the actual shape.

Our training data set consists of $4,000$ manually classified cells ($1,500$ each for slippers, and croissants, resp. and $1,000$ sheared croissants), which is augmented to the doubled number ($8,000$ in total) by mirroring each image along the centerline in flow direction. Each of those 4,000 initial training cell images represents one distinct cell, i.e. no cell was considered more than once in the training data set. We intentionally have fewer sheared croissants in the training data set than croissants and slippers since it is only an auxiliary cell class not reflected in the phase diagram but only to increase the precision of croissant classification. The three subsets (one per cell class) were taken at different pressure drops. Slippers were recorded at $700 - 1,000$ mbar, croissants at $100 - 200$ mbar and sheared croissants at $300 - 500$ mbar. Within these ranges, we applied several different pressure drops to ensure a certain variability of our training data set. All cells of a class are then mixed since they should be identified independent from the applied pressure. Further, we intentionally recorded cells for different lighting conditions by varying the power of the light source randomly (within a range of $50 - 80\,\%$). Together with a bilinear contrast adjustment of the resulting images (cf. experimental setup), this ensures high contrast dynamics of each image and corrects for minor changes in illumination. Due to constant manual observation during the recording process of the input data, we ensure collecting focused cell images. However, to ensure an invariance of our CNN approach with respect to optical misalignment and to reach a robust algorithm in terms of optical imaging, we also train for defocused cells by recording training images at different positions within a slightly tilted channel. The outcome will then be a neural network with a self-contained training set and thus is perfectly suited for a non-biased analysis.

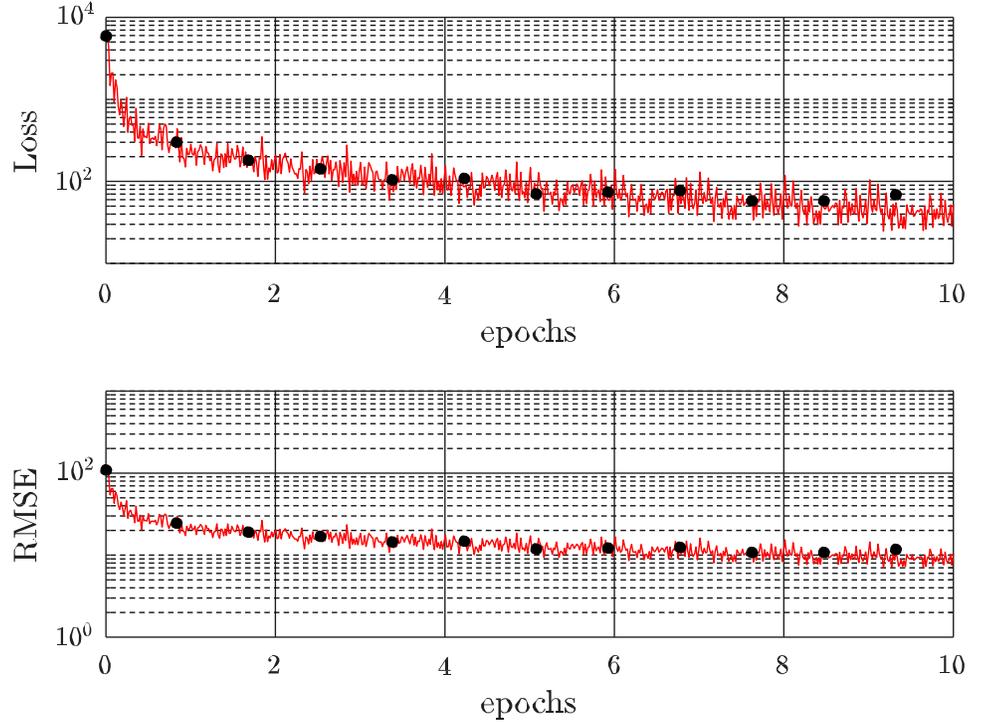

**Fig 3.** Diagram of training (red) and validation (black) status: Evolving convergence loss with growing number of training epochs. We set a maximum of ten epochs since a prolongation to more training epochs yields no gain in performance but rather causes overtraining. As training method, a gradient descent solver with momentum (SGDM) is used. The red line indicates the training loss, whereas the black dots represent the loss of the validation data set (validation loss). The progression of the validation loss serves as an indicator whether the CNN is overtrained, since the training and validation losses would diverge. As loss function, a root-mean-square error is chosen, being a standard approach for regression problems.

## Experimental setup

For our measurements, dilute suspensions of RBCs are required to observe single cells flowing through microcapillaries. Therefore, capillary blood is drawn with consent from a healthy individual and resuspended in a buffer solution (phosphate buffered saline, PBS, gibco by life technologies). After centrifugation at 1,500 $g$ for 5 minutes, RBCs sediment on the bottom of the sample tube, whereas leukocytes and platelets form a buffy coat on top of these (for further details concerning the blood sampling, see [30]). Since the presence of platelets can alter the dynamic properties of RBCs due to shear-induced platelet activation (SIPA, see [31]), this buffy coat is removed together with the supernatant and the residual pellet of RBCs is again resuspended in PBS. This procedure is repeated three times until the final pellet of cells will be adjusted to a hematocrit of $Ht \leq 1\%$ in a base solution of both PBS and bovine albumin (BSA, Sigma-Aldrich) at a concentration of 1 mg/ml, to avoid the well-known glass slide effect, turning discocytous RBCs to echinocytes.

With the aid of a high-precision pressure device (Elveflow OB 1 Mk II), this highly diluted suspension of RBCs is then driven through microcapillaries for a discrete set of 12 pressure drops $\Delta p \in \{20, 50, 100, 200, 300, 400, 500, 600, 700, 800, 900, 1000\}$ mbar, yielding various flow velocities, and recorded with a high-speed camera at a framerate of 400 Hz. Over the full pressure range, we record in total 3,090 RBCs and obtain experimental data, independent of the training data set.

The microcapillaries are formed in PDMS (polydimethylsiloxane), have a length of $L_x = 4$ cm and a rectangular cross-section with dimensions $L_y = 11.9 \pm 0.3\,\mu$m and $L_z = 9.7 \pm 0.3\,\mu$m. The $y$-direction is perpendicular to the camera-axis, whereas the $z$-direction points inward the camera axis. Several channels are branched from a common reservoir to the fluid inlet, which is connected to the pressure device via flexible tubing.

Post-processing the recorded media involves single particle tracking of cells, including distinction and sorting out non-isolated RBCs (i.e. RBCs being too close to each other, such that hydrodynamic interactions are not negligible). Due to the used optical setup (Nikon CFI Plan Fluor 60× oil-immersion objective, NA = 1.25) camera (Fastec HiSpec 2G) and capture settings, typical cell dimensions are in the range of 80 px. However, to allow for minor (physiological) variations, we crop the individual cell images to a format of $90 \times 90\,\text{px}^2$. Since the channel width is smaller than 90 px, we apply a Tukey window $w(y)$ (3) to the images, causing a smooth fade-out towards the channel walls

$$w(y) = \begin{cases} \frac{1}{2}\left[1 + \cos\left(\frac{2\pi}{\alpha}\left(y - \frac{\alpha}{2}\right)\right)\right], & 0 \leq y < \frac{\alpha}{2} \\ 1, & \frac{\alpha}{2} \leq y < 1 - \frac{\alpha}{2} \\ \frac{1}{2}\left[1 + \cos\left(\frac{2\pi}{\alpha}\left(y - 1 + \frac{\alpha}{2}\right)\right)\right], & 1 - \frac{\alpha}{2} \leq y \leq 1, \end{cases} \tag{3}$$

where $y$ denotes the relative position ($0 \leq y \leq 1$) in vertical direction of an arbitrary cell image. This reduces the influence of markable edges probably influencing the training and output of the CNN (Fig. 2).

For image preparation, we additionally map the image intensities to the full 8-bit range, such that the bottom 1 % and the top 1 % of all pixel values are saturated. This transformation yields higher signal dynamics and equal intensity profiles of all cell pictures and renders our approach more stable regarding slight illumination variations.

## Results and discussion

Neural networks are usually benchmarked by opposing their output with manually classified data. For our presented analysis, we compared the CNN-classified cell shapes with a manually ascertained phase diagram, presented first in [10]. This manual result serves as reference for all presented system benchmarks.

As can be seen in Fig. 4, the positions of the main populations of slippers (−117), croissants (115), and sheared croissants (≈ 40) differ significantly from the target values of −127, respectively 127 and 64. Deviations mostly occur due to different microscopy settings such as illumination and focusing. However, since thresholds are applied by statistical means, this shifting is not crucial for the final result. The center values for the occurring populations as well as the background are determined by applying four Gaussian fit functions.

Based on the standard deviations of the Gaussian distributions of the main classes, thresholds can be defined to assigning cells at certain position the respective population. Cells which do not fall in a designated range are classified as others rendering our regression based CNN an outlier tolerant classification system.

In Fig. 5, both the output values for the whole image data as well as the corresponding phase diagrams for different thresholds are shown. According to three predefined training classes (slippers, sheared croissants, croissants), we again fit the results of the CNN by four independent Gaussians, one of them accounting for the background.

In Fig. 5 (a), we labelled all cells within a confidence interval of $1\sigma$ ($1\sigma_{\text{croissant}}$, and $1\sigma_{\text{slipper}}$, resp.) to assign the cells to the associated distinct class. In Fig.

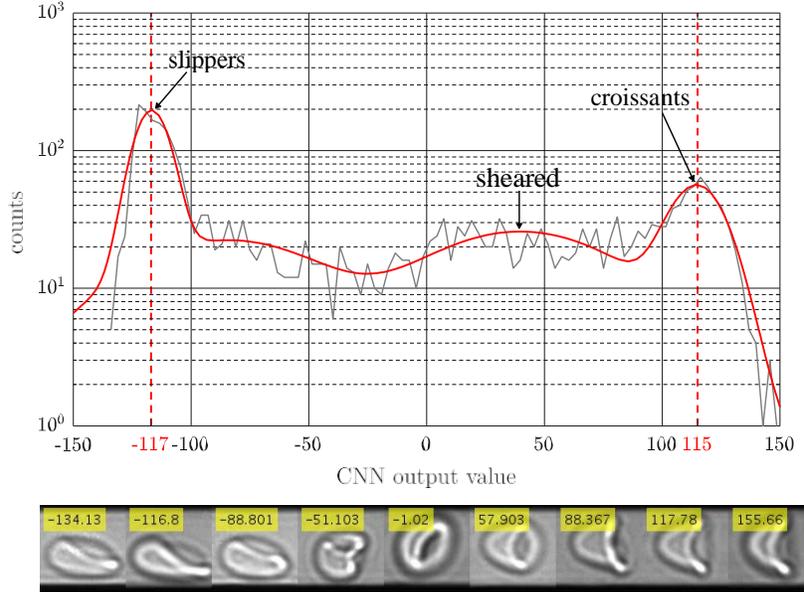

**Fig 4.** We estimate perfect slippers to be around the peak of the distribution at $\approx -117$, whereas croissants occur around $\approx 115$. By fitting the whole spectrum by four Gaussians, we are able to separate the respective contributions of each cell shape class and thus can determine a respective confidence interval. In the lower part, typical cell shapes are depicted for different output value ranges. Starting from the leftmost cell image, we undergo a shape change from slippers (image 1-3) to others (image 4-5) and finally to sheared (image 6-7) and pure croissants (image 8-9).

5 (b), manually evaluated data from [10] are depicted as areas, whereas the automatically evaluated data are marked as dash-dotted lines. The slipper population is represented in light blue, whereas croissant-labelled cells appear in light red. Although the qualitative behavior of the resulting phase diagram can be reproduced, for a threshold of $1\sigma$ we significantly underestimate the fraction of slippers and croissants compared to the result from manual classification.

By a change of these confidence bounds from $1\sigma$ to $2\sigma$, we perceive a better accordance of the manually classified croissants and the CNN-evaluated ones. Slippers still tend to be underestimated by the neural network, however, the deviation to the manually obtained graph is smaller throughout the whole velocity regime (Fig. 5 (c),(d)), causative of predicting a wrong phase transition point (intersection of both graphs). In Figs. 5 (e),(f), we therefore apply an adapted threshold range for each population, proposing a phase diagram in very good accordance to the benchmark of manually classified cell shapes. Even though a slight deviation to the manually generated phase diagram is still noticeable, both for slippers and croissants, the prediction of the phase transition point is in very good agreement with the original one, as well as the peaks of both cell fractions.

The underlying adapted confidence interval is derived from an iterative process, finding a minimum of a predefined cost function. As cost function, we calculate the number of false negatives and false positives for each cell shape class compared to the manual selection. By scanning the center values as well as threshold intervals of each population separately, both values (false positives and false negatives) are summarized to receive a cost value. Consequently, by minimizing this cost value, we obtain the optimal threshold range setting for each population. Finally, we yield adapted ranges at $-116.6 \pm 28.5$ for slippers and $116.8 \pm 35.5$ for croissants. All deviating individuals either for croissants and slippers are shown in Figs. 6 and 7. In total, the associated confusion matrix of our classification is depicted in Fig. 8.

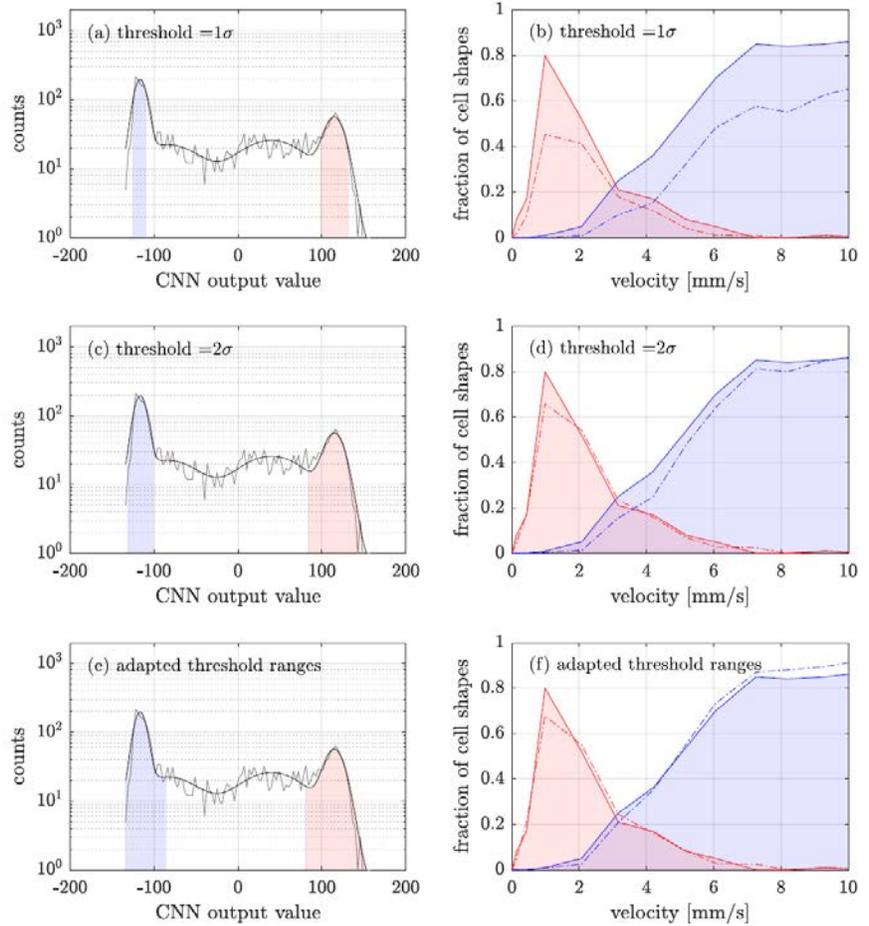

**Fig 5.** CNN output values for all cell images. The gray solid line is the network's output for the whole dataset, whereas the black solid line represents a fit with four Gaussians, one for each distinct class (croissants, slippers and sheared croissants, resp.) and one to account for indistinguishable cell shapes. The thresholds are shown in light blue and light red, respectively. In the right column, the obtained classification is compared with the manually ascertained phase diagram (solid lines). We stress the fact that the solid line is a guide to the eye, since we have a discrete number of flow veolcities due to the given number of applied pressure drops. In figure (b), a threshold of $1\sigma$ was used as a confidence interval to classify the cells into one of the two categories. Figures (d) and (f) show the resulting phase diagrams for a threshold of $2\sigma$ and an adapted $\sigma$, resp.

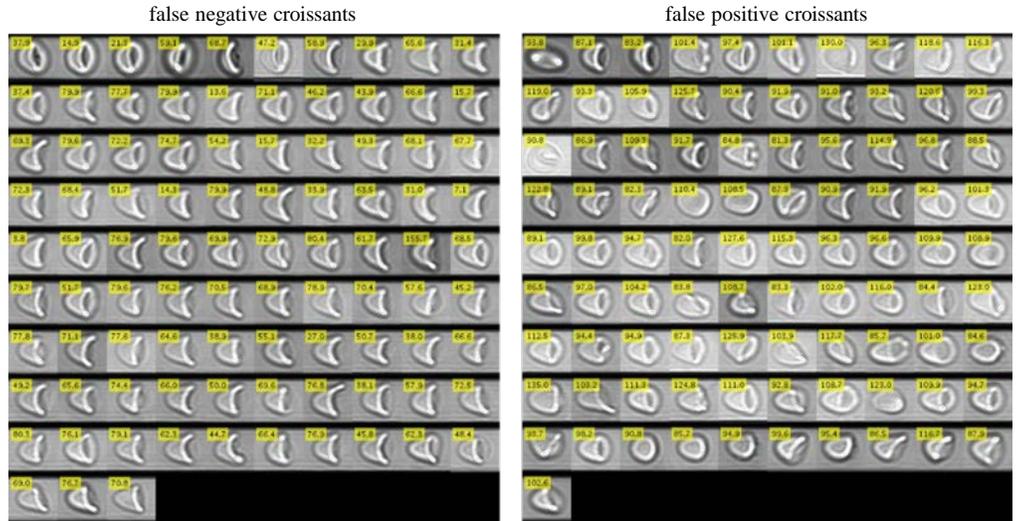

**Fig 6.** Image montage of all false negative (left) and false positive (right) classified croissants with respect to manual classification. On the left, false negative croissants are shown, i.e. all cells being classified as croissant manually, but not by the neural network. In contrast, all cells classified as croissant shapes by automated analysis but not by hand are depicted in the right montage (false positive croissants). Numerical values given in the yellow box of each picture correspond to the respective output value of the CNN.

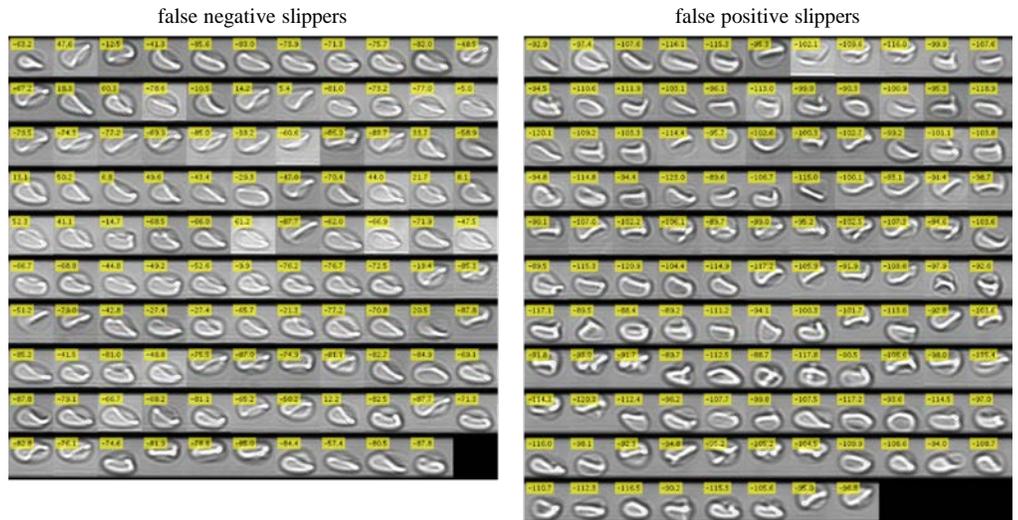

**Fig 7.** Image montage of all false negative (left) and false positive (right) classified slippers with respect to manually obtained classification. All cells being classified as croissants manually but not by the CNN (false negative slippers) are depicted in the left image, whereas all false positive slippers are shown in the right montage (cells classified as slipper shapes by automated analysis but not by hand). Additionally, each cell image contains a yellow box with the according CNN output values.

However, Figs. 6 and 7 illustrate that either the CNN as well as the manual evaluation of particular cells include many indifferent decisions especially for the false negative cases. Adjusted for these contrary decisions, it can be assumed that the false negative count rate influenced by subjective input is significantly lower. In contrast, with few exceptions, the false positive rate is in good accordance with the indicated value.

|  | actual class |  |  |
| --- | --- | --- | --- |
| predicted class | croissant | slipper | other |
| croissant | 551 (85.6%) | 0 (0.0%) | 91 (8.2%) |
| slipper | 0 (0.0%) | 1227 (91.8%) | 118 (10.6%) |
| other | 93 (14.4%) | 109 (8.2%) | 901 (81.2%) |

**Fig 8.** Confusion matrix with absolute values and relative percentages to evaluate the performance of the CNN approach. The rows hereby indicate the predicted, i.e. real class, whereas the columns indicate the actual class, corresponding to the CNN output. Thus, all values on the diagonal represent the correctly classified cells.

Referring to the final RMSE (Fig. 3) which corresponds to a value of $\approx 10$ regarding a target value of $|\pm 127|$, the false detection rates of slippers are in a very good agreement with the final error of the training process ($\approx 8\%$). For the croissants, we observe much higher deviations due to the occurrence of sheared individuals which are closely neighbored to the pure croissant population. This especially affects the false positive counts, whereas the false negative count contains a noticeably high amount of indifferent cases.

## Future perspectives

Using artificial intelligence for classification issues is a well-known, yet rare approach in biological systems. Although a wide variety of accessible, pre-trained, highly sophisticated neural networks (e.g. AlexNet, ResNet) exist, they tend to be over-engineered for most purposes in this field. Due to the complexity of their architecture with respect to the degrees of freedom, they require millions of input images. Furthermore, they are not designed for regression problems as required for the present study, but rather have a binary classification output. The presented CNN is a step forward to a fully automated analysis of recorded microscopy images associated with a big gain in evaluation efficiency. It moreover constitutes an unbiased classification system for RBC shapes in flow. Any restriction to cell classes is purely artificial in the sense that we train the CNN solely with RBC shapes in flow. Similarly, one could tailor a set of training data for other distinct cell classes or even cell types. We aim to conduct further experiments with different channel geometries and flow conditions, amending the existing phase diagram.The phase diagram and especially the phase transition point between croissants and slippers can be altered adding drugs to the RBC suspension. Since certain drugs, e.g. acetylsalicylic acid, show a strong effect on the membrane structure [32], we conjecture to resolve this feature in phase diagrams of future studies, leading to insights towards mechanical properties of individual RBCs in flow. Following this scope, the change in flow behavior of whole blood under drug influence can be predicted due to the influence on its main constituent, the red blood cells. This might play a key role in finding appropriate drugs to avoid pathogenic incidents, e.g. stenosis. Analogously, we clearly see evidence to detect maladies causing an alteration of cell's intrinsic mechanical properties, e.g. in sickle cell disease, thalassemias and numerous rare anaemias. By slight technical adaptations (e.g. the usage of high-performance graphics cards), it is also possible to classify cells on-the-fly, i.e. in real-time rather than with the presented frame-based approach. Prospectively, the here proposed CNN will be a useful quantitative tool in hematology aiming to investigate cell membrane characteristics.